\documentclass{PoS}
\usepackage{amsmath}
\title{The physics of strong magnetic fields  and activity of magnetars}
\ShortTitle{The physics of strong magnetic fields  and activity of
magnetars}
\author{\speaker{Qiu He Peng}~ and Hao Tong\\
Department of Astronomy,
 Nanjing University, \\Nanjing, 210093, China\\ E-mail: \email{qhpeng@nju.edu.cn}}

\abstract{A phase transition from paramagnetism to ferromagnetism in
neutron star interior is explored. Since there is $^3$P$_2$ neutron
superfluid in neutron star interior, it can be treated as a system
of magnetic dipoles. Under the presence of background magnetic
field, the magnetic dipoles tend to align in the same direction.
Below a critical temperature, there is a phase transition from
paramagnetism to ferromagnetism. And this gives a convenient
explanation of the strong magnetic field of magnetars. In our point
of view, there is an upper limit for the magnetic field strength of
magnetars. The maximum field strength of magnetars is about
$(3.0-4.0)\times 10^{15}$ G. This can be tested directly by further
investigations.

Magnetars are instable due to the ultra high Fermi energy of
electrons. The Landau column becomes a very long cylinder along the
magnetic field, but it is very narrow and the Fermi energy of
electron gas is given as $E_F(e)\approx40(B/B_{cr})^{1/4}$ when
$B\gg B_{cr}$. $E_F(e)\approx90MeV$ When $B\sim10^{15}$ G. Hence,
the electron capture process $e^-+p\to n+\nu_e$ will be happen
rapidly. Thus the $^3$P$_2$ Cooper pairs will be destroyed quickly
by the outgoing neutrons with high energy. It will cause the
isotropic superfluid disappear and then the magnetic field induced
by the $^3$P$_2$ Cooper pairs will be also disappear. These energy
will immediately be transmitted into thermal energy and then
transformed into the radiation energy with X-ray - soft
$\gamma$-ray. We may get a conclusion that the activity of magnetars
originates from instability caused by the high Fermi energy of
electrons in extra strong magnetic field. }

\FullConference{10$^{\rm th}$ Symposium on Nuclei in the Cosmos
 \\27 July - 1 August 2008\\Mackinac Island,Michigan, USA }

\begin{document}
\section{Introduction}
\vspace*{-0.2cm}A puzzle is what is the origin of strong magnetic
field of magnetars and it is still open question, although it has
been investigated by many authors.

In this paper we propose a new idea for the origin of the magnetars.
The strong magnetic fields of the magnetars may originate from the
induced magnetic moment of the $^3$P$_2$ neutron Cooper pairs in the
anisotropic neutron superfluid.

\section{Induced paramagnetic moment of the $^3$P$_2$ neutron superfluid and the Upper limit of the magnetic field of magnetars}
\vspace*{-0.2cm} A magnetic moment of the $^3$P$_2$ neutron Cooper
pair  is twice that of the abnormal magnetic moment of a
neutron,$2\mu_n$ in magnitude, and its projection on the external
magnetic field (z-direction) is $\sigma \times (2\mu_n)$ , $\sigma_z
=1,0,-1$, where $\mu_n$ is the absolute value of the magnetic moment
of a neutron, $\mu_n=-0.966 \times 10^{-23}$ erg/G.

A magnetic dipole tends to align in the direction of the external
magnetic field. The $^3$P$_2$ neutron Cooper pair has energy
$\sigma_Z\times\mu_nB$ in the applied magnetic field due to the
abnormal magnetic moment of the neutrons. B is the total magnetic
field including the background one.

The difference of the number density of $^3$P$_2$ neutron Cooper
pairs with paramagnetic and diamagnetic moment is
\begin{equation}\label{eq23}
\Delta n_{\mp}=n_{-1}-n_{+1}=n_n(^3P_2)f(\frac{\mu_nB}{kT})
\end{equation}
\begin{equation}\label{eq24}
f(x)=\frac{2\sinh(2x)}{1+2\cosh(2x)}
\end{equation}
The Brillouin function, $f(\mu_n B/kT)$, is introduced to take into
account the effect of thermal motion We note that $f$  is an
increasing function, in particular,$f(x)\approx 4x/3$, for $x\ll 1$
and $f(x)\to 1$, when $x\gg 1$. $f(\mu_nB/kT)$ increase with
decreasing temperature. And this is the mathematical formula for the
B-phase of the $^3$P$_2$ superfluid.

A relevant question is how many neutrons have been combined into the
$^3$P$_2$ Cooper pairs?  The total number of neutrons is given by
$N=2V\int^{k_F}_0\frac{d^3k}{(2\pi)^3}=\frac{V}{3\pi^2}k^3_F$ (Here
$k$ is a wave vector). The neutrons combined into the $^3$P$_2$
Cooper pairs are just in a thin layer in the Fermi surface with
thickness $k_{\Delta}$\cite{Lifshitz}, $\hbar k_{\Delta}=\sqrt{2m_n
\Delta_n(^3P_2)}$ (We would like to emphasize that the energy gap,
$\Delta$, is the binding energy of the Cooper pair rather than a
variation of the Fermi energy due to the variation of particle
number density). Then we have $\delta N \approx
\frac{V}{\pi^2}k_{f}^{3}\frac{k_{\Delta}}{k_F}(k_{\Delta}<<k_F)$.
Thus, the fraction of the neutrons that combined into the $^3$P$_2$
Cooper pairs is
\begin{equation}\label{eq25}
q=\frac{\delta N}{N}\approx 3\frac{k_{\Delta}}{k_F}
\end{equation}

The fraction of the neutrons that combined into the $^3$P$_2$ Cooper
pairs is ($p=\hbar k$)
\begin{equation}\label{eq26}
q=\frac{4\pi
p^2_F[2m_n\Delta(^3P_2)]^{1/2}}{(4\pi/3)p^3_F}=3[\frac{\Delta
(^3P_2)}{E_F}]^{1/2}.
\end{equation}

The Fermi energy of the neutron system is $E_F \approx 60
(\frac{\rho} {\rho_{nuc}})^{2/3}$ Mev. The energy gap of the
anisotropic neutron superfluid is $\Delta(^3P_2) \sim 0.05$ MeV
\cite{Elgag}, $q \sim 8.7\%$.
 Thus, the total number of the $^3$P$_2$ Cooper pairs is $n_n(^3P_2)\approx q N_A m(^3P_2)/2$ .
 Therefore, the total difference of the $^3$P$_2$ neutron Cooper
 pair number with paramagnetic and diamagnetic moment is
\begin{equation}\label{eq27}
\Delta
N_{\mp}=n_n(^3P_2)f(\frac{\mu_nB}{kT})=\frac{1}{2}N_Am(^3P_2)q
f(\frac{\mu B}{kT})
\end{equation}
The total induced magnetic moment, of the anisotropic neutron
superfluid is
\begin{equation}\label{eq28}
\mu^{(in)}(^3P_2)=2 \mu_n \times \Delta N_{\mp}=\mu_n N_A m(^3P_2)q
f(\mu_nB/kT)
\end{equation}
Where $m(^3$P$_2)$  is the mass of the anisotropic neutron
superfluid in the neutron star, $N_A$ is the Avogadro constant. The
induced magnetic moment is just the fully magnetized quantity $\mu_n
N_A m(^3P_2)$, with two modification factors. The factor $q$ takes
into account the Fermi surface effect. While $f(\mu_nB/kT)$ is the
thermal factor taking into consideration the finite temperature
effect. For a dipolar magnetic field \cite{Shapiro}
$|\mu_{NS}|=B_{\gamma}R^3_{NS}/2$. Here $B_p$ is the polar magnetic
field strength and $R_{NS}$ is the radius of the neutron star. The
induced magnetic field is then
\begin{equation}\label{eq29}
B^{(in)}=\frac {2 \mu^{(in)}(^3 P_2)}{R^3_{NS}}=\frac {2 \mu_n
N_A(^3P_2)}{R^3_{NS}}q f(\mu_nB/kT)
\end{equation}

\begin{equation}\label{eq31}
B^{(in)}_{max}(^3P_2)=\frac {2 \mu_n N_A(^3P_2)}{R^3_{NS}}q \,\,\,\,
G\approx 2.02\times 10^{14}\eta \,\,\,\,G
\end{equation}

\begin{equation}\label{eta}
\eta=\frac{m(^3P_2)}{0.1M_{sun}}R^{-3}_{NS,6}[\frac{\Delta_n(^3P_2)}{0.05MeV}]^{1/2}
\end{equation}

Here $\eta$ is the dimensionless factor describes both the
macroscopic and microscopic properties of neutron stars.

We note that the induced magnetic field for the anisotropic neutron
superfluid increases with decreasing temperature as the Brillouin
function $f(\mu_nB/kT)$ which tends to 1 when the temperature
decreases low enough. Actually, $f(\mu_nB/kT)\sim1$ as long as
$\mu_nB/kT \gg 1$. For example, this is true when $T\leq10^7$K if
$B= 10^{15}$ G.

There is an upper limit for the induced magnetic field of the
$^3$P$_2$ superfluid according to eq.(\ref{eq31}). It corresponds to
the maximum value unity of the temperature factor $f(\mu_nB/kT)$.
This upper limit can be realized when all the magnetic moments of
the $^3$P$_2$ neutron Cooper pairs are arranged with the
paramagnetic direction as the temperature become low enough.

The maximum magnetic field for magnetars depends on the total mass
of the anisotropic neutron superfluid of the neutron star. The upper
limit of the mass for the neutron stars is more than
$2M_{sun}$\cite{Lattimer} . It is therefore possible that the mass
of the anisotropic neutron superfluid of the heaviest neutron star
may be about $(1-1.5)M_{sun}$. Hence, the maximal magnetic field for
the heaviest magnetar may be estimated to be
$(3.0-4.0)\times10^{15}$ G. This can be tested directly by magnetar
observations.

\section{From paramagnetism to ferromagnetism}

\vspace*{-0.2cm}We have $\mu_nB\ll kT$ when $B<10^{13}$ G and $T
> 3\times 10^6$ K, therefore we could make Taylor series
\begin{equation}\label{eq41}
B^{(in)}\approx
\frac{8}{3}\frac{\mu_nN_Am(^3P_2)}{R^3_{NS}}q\frac{\mu_nB}{kT}\approx
\frac{1.9}{T_7}\eta B
\end{equation}
\begin{equation}\label{eq42}
B=B^{(in)}+B^{(0)}
\end{equation}
Where $B^{(0)}$ is the applied magnetic field which includes both
the initial magnetic field of the collapsed supernova core and the
induced magnetic field produced by the Pauli paramagnetization of
the highly degenerate relativistic electron gas in the neutron star
interior \cite{Peng2}.

We may solve the induced magnetic field by combining eq.(\ref{eq41})
and eq.(\ref{eq42}):
 \begin{equation}\label{eq43}
B^{(in)}=(\frac{T_7}{1.9\eta}-1)^{-1}B^{(0)}
\end{equation}
The induced magnetic field, $B^{(in)}$ , of the anisotropic neutron
superfluid is much weaker than the applied magnetic field ,
$B^{(0)}$, when the temperature is very high, $T_9\gg 1.9\eta$ .
However, the induced magnetic field of the anisotropic neutron
superfluid would exceed the applied magnetic field when the
temperature decreases down to $1<T_7/(1.9\eta)<2$. Therefore
$B^{(in)}$ should be calculated by both eq.(\ref{eq31}) and
eq.(\ref{eq42}) when $T_7\sim 1.9\eta$. This belongs to the domain
of ferromagnetism (see \cite{Feng}\cite{Pathria}).

In view of the above discussions we may get a very important
conclusion: the strong magnetic field of magnetars may originate
from the induced magnetic field by the ferromagnetic moments of the
$^3$P$_2$ Cooper pairs of the anisotropic neutron superfluid at a
moderate low temperature about $10^7$K.

\section{Activity of Magnetars}
\vspace*{-0.2cm}The core temperature of the magnetars is about
$10^7$ K in our model. While observations show that some SGR's and
AXP's have high thermal-type-spectrum X-ray flux, being among the
hottest neutron stars.  We discuss this question and give reasonable
and consistent explanation in this paragraph.

Energy of an electron under the super strong magnetic field is
quantized. It is
\begin{equation}\label{eq51}
E^2_e(p_z,B,n,\sigma)=m^2_ec^4+p^2_zc^2+(2n+1+\sigma)2m_ec^2\mu_eB
\end{equation}
(The magnetic moment of an electron is $0.927¡Á10^{-20} erg/G$) The
overwhelming majority of electrons congregate in the lowest levels
$n=0$ or $n=1...$ when $B\gg B_{cr}$. The Landau column is a very
long cylinder along the magnetic field and it is very narrow.

We may find the Fermi energy of the electrons by solving the
following complicated integral equation

\begin{equation}\label{eq52}
N_{total}=N_AY_e\rho=\frac{2\pi}{h^3}\int^{p_F}_{0}dp_z\sum_{\sigma=\pm
1}\int\delta((p_{\bot}/m_ec)-[(2n+1+\sigma)b]^{1/2})p_{\bot}dp_{\bot}
\end{equation}
\begin{equation}\label{eq53}
n_{max}(p_z,b,\sigma)=Int\{\frac{1}{2b}[(\frac{E_F}{m_ec^2})^2-1-(\frac{p_z}{m_ec})^2]-1-\sigma\}
\end{equation}
\begin{equation}\label{eq54}
b=B/B_{cr}
\end{equation}
$N_{total}$ is the total occupied number of electrons in a unit
volume and Int\{...\} is the integer function. After some
calculation, we may get the approximation expression
\begin{equation}\label{eq55}
E_F(e)\approx40(B/B_{cr})^{1/4}~~~~~~~~\text{when}~~ B\gg B_{cr}
\end{equation}
We have $E_F(e)\approx 90MeV$, when B $\sim10^{15} G$.

Magnetars are instable due to the ultra high Fermi energy of
electrons. Hence, the electron capture process $e^-+p\to n+\nu_e$
will be happen rapidly as long as the Fermi energy of electrons is
much greater than 60 Mev which is the Fermi energy of neutrons.
Energy of the resulting neutrons will be rather high and they will
react with the neutrons in the $^3$P$_2$ Cooper pairs and thus the
$^3$P$_2$ Cooper pairs will be destroyed quickly by the process
$n+(n\uparrow,n\uparrow)\rightarrow n+n+n $. It will cause the
isotropic superfluid disappear and then the magnetic field induced
by the $^3$P$_2$ Cooper pairs will be also disappear.

The remaining energy per outgoing neutron after the process above
will be approximately
\begin{equation}
\bar{\epsilon}\approx
\frac{1}{3}[E_F(e)+E_F(p)-E_F(n)-(m_n-m_p-M_e)c^2]
\end{equation}
These energy will immediately be transmitted into thermal energy and
then transformed into the radiation energy with X-ray - soft
$\gamma$-ray.

The total released energy may be approximately estimated as
\begin{equation}\label{eq58}
E^{(tot)}\approx\bar{\epsilon}\times n(^3P_2)= 3\bar{\epsilon}\times
qN_Am(^3P_2)\\
\approx2.0\times
10^{51}[(\frac{B}{B_{cr}})^{1/4}-1.5]\frac{m(^3P_2)}{0.1M_{sun}}
ergs
\end{equation}

The x-ray Luminosity of AXPs is $L_x\approx
10^{34}-10^{36}\,\,erg/sec$. It will be enough to maintain the
luminosity of AXPs over $10^8 $yr.

\textbf{We may get a conclusion that the activity of magnetars
originates from instability caused by the high Fermi energy of
electrons in extra strong magnetic field.} However, we have to
calculate rate of the electron capture process for given magnetic
field to see if our idea is reasonable for the x-ray luminosity of
AXPs and their surface temperature. This is our next work.

\acknowledgments{The author Q-h Peng is very grateful to Prof.
Chich-gang Chou for his help of improving the presentation of the
paper. This research is supported by Chinese National Science
Foundation through grant no.10573011, grant no.10273006, and grant
no. 10773005 and the Doctoral Program Foundation of State Education
Commission of China.}

\end{document}